# Quantitative analysis of image quality in low-dose CT imaging for Covid-19 patients

**Behrooz Ghane,[a] Alireza Karimian,[b] Samaneh Mostafapour,[c] Faezeh Gholamiankhak,[d] Seyedjafar Shojaerazavi,[e] Hossein Arabi[f]**

[a]University of Isfahan, Faculty of Engineering, Department of Biomedical Engineering, Isfahan, Iran

[b]University of Isfahan, Faculty of Engineering, Department of Biomedical Engineering, Isfahan, Iran

[c]Mashhad University of Medical Sciences, Faculty of Paramedical Sciences, Department of Radiology Technology, Mashhad, Iran

[d] Shahid Sadoughi University of Medical Sciences, Faculty of Medicine, Department of Medical Physics, Yazd, Iran

[e]Mashhad University of Medical Sciences, Ghaem Hospital Mashhad, Department of Cardiology,Mashhad, Iran

[f] Geneva University Hospital, Division of Nuclear Medicine and Molecular Imaging, CH-1211 Geneva 4, Switzerland

**Abstract**

**Purpose**: Immediate and accurate diagnosis of COVID-19 helps to prevent its irreversible impacts and rapid outbreak. CT scan together, with the real-time reverse transcription-polymerase chain reaction (RT-PCR) test are the main tools to diagnose and grade COVID-19 progression. To avoid the side-effects of CT imaging, low-dose CT imaging is of crucial importance to reduce population absorbed dose. However, this approach introduces considerable noise levels in CT images, which might have an adverse impact on the diagnosis and grading of COVID-19.

**Approach**: In this light, we set out to simulate four reduced dose-levels (60%-dose, 40%-dose, 20%-dose, and 10%-dose) of standard CT imaging using Beer-Lambert's law across 49 patients infected with COVID-19. Then, three denoising filters, namely Gaussian, Bilateral, and Median, were applied to the different low-dose CT images, the quality of which was assessed prior to and after the application of the various filters via calculation of peak signal-to-noise ratio (PSNR), root mean square error (RMSE), structural similarity index measure (SSIM), and relative CT-value bias, separately for the lung tissue and whole-body.

**Results**: The quantitative evaluation indicated that 10%-dose CT images have inferior quality (with RMSE=322.1±104.0 HU and bias=11.44±4.49% in the lung) even after the application of the denoising filters. The bilateral filter exhibited superior performance to suppress the noise and recover the underlying signals in low-dose CT images compared to the other denoising techniques. The bilateral filter led to RMSE and bias of 100.21±16.47 HU, -0.21±1.20%, respectively in the lung regions for 20%-dose CT images compared to the Gaussian filter with RMSE=103.46±15.70 HU and bias=1.02±1.68%, median filter with RMSE=129.60±18.09 HU and bias=-6.15±2.24%, and the nonfiltered 20%-dose CT with RMSE=217.37±64.66 HU and bias=4.30±1.85%.

**Conclusion**: In conclusion, the 20%-dose CT imaging followed by the bilateral filtering introduced a reasonable compromise between image quality and patient dose reduction.

# 1 Introduction

The World Health Organization (WHO) identified Coronavirus 2 (SARS-CoV-2), which causes a severe acute respiratory syndrome, as a global public health concern.[1] Up to this date (January 30, 2021), more than 100 million cases have been diagnosed with Coronavirus disease.[2] Coronavirus or Covid-19 affects the respiratory system and causes mild to severe respiratory syndromes, including pneumonia. The management, outbreak prevention, and treatment of this disease are currently the most critical global challenges in medicine.[3,4] Therefore, an accurate and quick diagnosis of this disease is of essential importance. Real-time reverse transcription-polymerase chain reaction (RT-PCR) test is regarded as a standard method for the diagnosis of Covid-19. However, this method has severe limitations such as the preparation time, considerable false positive and false negative rates[5], and relatively low sensitivity[6]. Preliminary studies have confirmed that computed tomography (CT) imaging is an effective approach for diagnosing Covid-19 disease.[7,8] CT imaging is an accurate and non-invasive method for detecting abnormalities, including tumors, bone fractures, and cardiovascular diseases.[9-12] Recently, an extended range of clinical studies has been performed on the potentials of CT imaging for the diagnosis and management of patients with Covid-19 disease. These studies, in fact, continue to fully identify the underlying Covid-19 features in CT images.[8]

Although CT imaging is a rapid method for the diagnosis of Covid-19 disease, the absorbed radiation dose by patients, due to the x-ray exposure, is one of its major concerns. The radiation risks of CT scans are seriously high for those required to undergo multiple scans, pregnant women, and pediatrics.[13] A recent study showed that DNA double-strand breaks and chromosome aberrations would increase in patients who undergo standard-dose CT scans, while there was no DNA damage in people who have been examined with low-dose CT scans.[14] Despite the significant advances made in software and hardware technology of CT scanners[15,16], it is not yet considered as a low-dose safe imaging modality[17]. Therefore, regarding CT scan's side-effects[18], the application of low-dose CT protocols would be of crucial importance in routine CT imaging. However, increased image noise is a major limitation in this imaging modality.[19]

The diagnosis of Covid-19 (infection) patterns within the lung region would be highly challenging in CT images with increased noise levels, which can skew the correct diagnosis. Reducing noise levels in CT images and enhancing image quality has always been one of the important research topics.[20]

Noise reduction in low dose CT imaging can be achieved in three ways[21]: denoising in the sinogram or projection space prior to the image reconstruction[22,23], using the dedicated reconstruction algorithms which model the noise within image formation[24,25], and denoising in the image domain using post-reconstruction algorithms[19,26]. The induced-noise in low-dose CT images normally follows a Poisson noise distribution, Gaussian noise distribution, or a combination of the two.[27] The commonly used denoising techniques in medical imaging are Gaussian, median, and bilateral filters which could be employed either in the projection or image space.[28,29] Implementation of these techniques in the projection space requires access to the raw acquisition data which is not commonly provided by CT scanners.[21] Iterative CT image reconstruction algorithms tend to model the underlying causes of noise within the image reconstruction process to suppress the noise in the resulting images, commonly through the maximum likelihood formulation of the Poisson noise. These techniques, in addition to high-computation time, also require raw acquisition data. Moreover, they should be dedicatedly developed/optimized for each scanner and acquisition protocol to properly perform noise suppression.[21,30,31]

The denoising techniques in the image domain, contrary to the other approaches, do not require raw sinogram data and can directly be applied to CT images. In addition to low computation time, these techniques are almost independent of CT scanners, acquisition protocols, and image reconstruction algorithms. Moreover, they could easily be adapted to the new acquisition protocols and, or noise levels through the adjustment of a few hyperparameters.

In addition to the above-mentioned approaches, deep learning-based denoising techniques have exhibited promising performance in low-dose CT imaging.[32,33] These methods require an extensive training dataset to create a robust model to inter-subject image quality and, or anatomical variations. Moreover, these techniques are very sensitive to abnormal and, or odd cases wherein they may lead to gross errors and misdiagnosis.[19,34,35]

The aim of this work is to investigate low-dose CT imaging for Covid-19 patients, and in particular, how reduced patient dose would affect the quality of the CT images. To this end, the low-dose CT imaging of Covid-19 patients will be simulated for different dose levels up to 10% of the standard dose to measure the increased noise levels and quantitative bias. Thereafter, different post-reconstruction denoising approaches, namely Gaussian, median, and bilateral filters, will be applied to the low-dose CT images to suppress the noise and recover the image quality. The goal is to determine which reduced dose level and post-reconstruction denoising approach would lead to an optimal compromise between image quality and patient radiation dose.

## 2 Material and method

### 2.1 Data Acquisition

Chest CT scans from 49 patients with positive Covid-19 (approved by the RT-PCR test) were acquired between January and March of 2020 in Shariati Hospital, Mashhad, Iran. CT image acquisitions were then performed on a Siemens Somatom Spirit Dual Slice CT with tube energy of 130 kVp, tube current of 48 mAs, CT Dose Index (CTDI$_{VOL}$) of 5.64 mGy, Dose Length Product (DLP) of 132.32 mGy.cm, rotation time (TI) of 0.8 s, and slice thickness of 5 mm. For the entire group of patients, the lung region was semi-automatically segmented from the standard CT images using Pulmonary Toolkit (PTK) software.[36] The lung segmentation was visually verified and the miss-segmentation errors were manually corrected.

*2.2 Low-dose CT simulation*

In this study, low-dose CT imaging was simulated for reduced doses of 60%, 40%, 20%, and 10% of the standard CT images (reference). To this end, the method proposed by[37] was employed to simulate low-dose CT imaging as described in the following steps.

The parallel beam geometry estimates the spatial coordinates of the x-ray beam passing through the subject in the field of view. Figure 1 depicts the concept of parallel beam geometry and its key parameters.

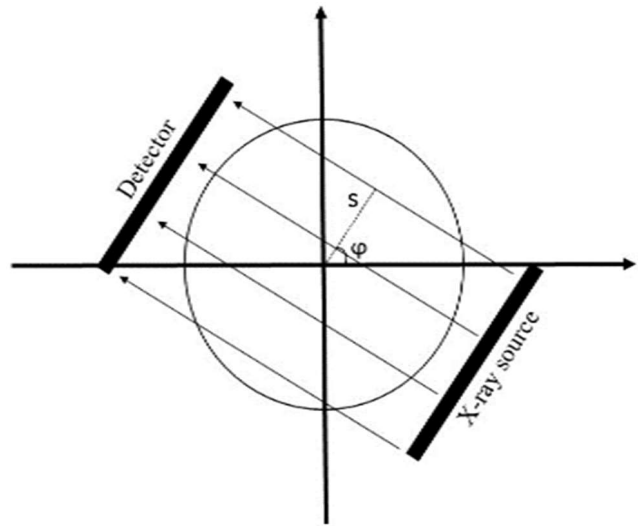

**Fig. 1** Parallel beam geometry and its key parameters.

A single point in space is defined by a combination of $\varphi$ and $s$ parameters which are the angle between the x-ray beam and the distance of the x-ray beam from the center, respectively. Since the $L_{s, \varphi}$ line is perpendicular to the x-ray beam, it can be described by a point $(x, y)$ on $L_{s, \varphi}$ using a real number $t \in \mathbb{R}$ and $\varphi \in (0, 2\pi)$ as follows:

$$L_{s, \varphi}(t) = (x(t), y(t)) = s(\cos\varphi, \sin\varphi) + t(-\sin\varphi, \cos\varphi). \qquad (1)$$

Then, function $f(L_{s, \varphi})$ can be defined on $\mathbb{R}^2$ as:

$$Rf(s, \varphi) = \int_{-\infty}^{+\infty} f(L_{s, \varphi}(t)) dt. \qquad (2)$$

In the above equation, $R$ is the Radon transform which is applied to the $f$ function.

Based on Beer-Lambert's law, the result of applying the Radon transforms in Eq. (2) would be

$$Rf(s, \varphi) = -\ln\left(\frac{I_1(s,\varphi)}{I_0(s,\varphi)}\right) = y(s, \varphi) \qquad (3)$$

$I_0$ is the intensity at the x-ray source and $I_1$ is the intensity (signal) recorded in the detector.

According to Eq. (3), the added noise on the CT images could be modeled by a Poisson distribution as follows:

1. Eq. (4) calculates Hounsfield Unit (HU) values (CT numbers) from the standard-dose CT images (reference), where *Intercept* and *Slope* values are obtained from DICOM header information.

$$HU_s = (PixelValue \times Slope) + Intercept \qquad (4)$$

2. Considering the tube voltage in standard CT imaging, we assume $\mu_{air}=0$ and the linear attenuation coefficients were computed using Eq. (5)

$$\mu_{tissue} = \left(HU_s \times \frac{\mu_{water}}{1000}\right) + \mu_{water} \qquad (5)$$

3. The projection data ($\rho_s$) is obtained from the application of the Radon transform to the reference CT images (converted to the linear attenuation coefficients).
4. The normal-dose transmission data ($T_{nd}$) is calculated using Eq. (6).

$$T_{nd} = exp\ (-\rho_s) \quad (6)$$

5. The low-dose transmission data $(T_{ld})$ is generated by adding Poisson noise to the original data

$$T_{ld} = Poisson\ (I_0 \times T_{nd}), \quad (7)$$

where $I_0$ is the incident flux for simulating low-dose scans.

6. The low-dose projection data $(\rho_{ld})$ is calculated in the sinogram domain using Eq. (8).

$$\rho_{ld} = ln\ \left(\frac{I_0}{T_{ld}}\right) \quad (8)$$

7. Finally, the inverse of Eq. (5) is applied to convert the attenuation coefficients to the CT values and the final low-dose image is reconstructed using the inverse Radon transform and filter back-projection algorithm.

Based on Beer-Lambert's law, the intensity index $I_0$ indicates the number of photons per detector bin in the air scan, which is chosen 6000 for 60% low-dose simulation. According to Ref. 37, $I_0$ was assumed $10^4$ in the full-dose imaging, thus $I_0 = 6000$ was considered as 60% low-dose, $I_0 = 4000$ as 40%, $I_0 = 2000$ as 20%, and $I_0 = 1000$ as 10% low-dose CT imaging.

## 2.3 Denoising algorithms

### 2.3.1 Gaussian Filter

Gaussian filter is a simple and intuitive approach of noise suppression that replaces the value of each pixel with the weighted average of surrounding ones. This approach may introduce noticeable signal loss and blur in the CT images.[29] The Gaussian filter is defined as

$$F(x, y) = \frac{1}{2\pi\sigma^2} e^{-\frac{x^2+y^2}{2\sigma^2}} \quad (9)$$

where, *x* and *y* are the vertical and horizontal distances from the target pixel (in two-dimensional implementation) and σ is the standard deviation of the Gaussian distribution. The kernel obtained from this equation is convolved with the noisy CT images to diminish the noise levels. The standard deviation of the Gaussian kernel is the key parameter that defines the levels of smoothness in the resulting images. The procedure to find optimal standard deviation values for the different noise levels (low-dose CT images) is described in the following section.

*2.3.2 Median Filter*

The median filter, as a non-linear denoising approach, is commonly used for noise suppression in natural as well as medical images. The median filter is functionally similar to the moving average filter, but it calculates the median value of the pixels/voxels rather than the mean value. Median filters might exhibit better performance regarding the edge preservation in the filtered images than the Gaussian filter. The median filter is formulated as

$$F(x, y) = median(x, y). \tag{10}$$

where $x \times y$ defines the size of the filter window for calculation of the median value. The filter window size should be optimized independently for the different low-dose CT images.

*2.3.3 Bilateral Filter*

Bilateral filter, regarded as an edge-preserving smoothing approach, tends to suppress noise while maintaining the prominent edges/patterns of the image through penalizing the smoothing function over the edges.[38] The Bilateral filter, employed to reduce the noise levels in the different low-dose CT images, is mathematically formulated as

$$w(i,j,k,l)=exp(-\frac{(i-k)^2+(j-l)^2}{2\sigma_d^2}-\frac{||I(i.j)-I(k.l)||^2}{2\sigma_r^2}) \tag{11}$$

Here, *(i,j)* and *(k,l)* indicate the location of the target voxel to be filtered as well as its neighboring voxels. *w(i,j,k,l)* returns the weighting factors for each of the neighboring voxels

*(k,l)* based on spatial closeness (first term in Eq. 11) and intensity difference (second term in Eq. 11) with respect to the target voxel *(i,j)*. *I(i,j)* and *I(k,l)* denote the intensity of *(i,j)* and *(k,l)* voxels, respectively, and $\sigma_d$ and $\sigma_r$ are the smoothing parameters. The calculated weights from Eq. 11 are used in Eq.12 to estimate the noise-free voxel values ($I_D$) in the filtered images.

$$I_D (i,j) = \frac{\sum_{k,l} I(k,l) w(i,j,k,l)}{\sum_{k,l} w(i,j,k,l)} \quad (12)$$

## 2.4 Quantitative evaluation

The performance of the Gaussian, Median, and Bilateral denoising techniques was evaluated on the CT images with different dose levels using the following four image-quality metrics.

Peak Signal-to-Noise Ratio (PSNR) presents the ratio between the maximum intensity of the noisy or reference image, and the mean square error (*MSE*), which is usually expressed in the unit of logarithmic decibel as follows:

$$PSNR(y, x) = 10 \log_{10} \left( \frac{\max(x)^2}{MSE(y.x)} \right) \quad (13)$$

Here, *x* is the full-dose CT image (reference), *y* is the low-dose (or filtered) CT image and *MSE (y, x)* stands for the mean squared error between the reference and low-dose (filtered) CT images as follows:

$$MSE = \frac{1}{mn} \sum_{i=0}^{m-1} \sum_{j=0}^{n-1} [x(i,j) - y(i,j)]^2 \quad (14)$$

Here, *m×n* (row×column) is the number of voxels in the CT images and *i* and *j* stand for the voxel indices.

Structural Similarity Index Measure (SSIM), which reflects the perceived quality of the images, was used to compare the overall structural similarity between low-dose (or filtered) and the reference CT images.

$$SSIM\ (x,\ y) = \frac{(2\mu_x\mu_y+c_1)(2\sigma_{xy}+c_2)}{(\mu_x^2+\mu_y^2+c_1)(\sigma_x^2+\sigma_y^2+c_2)} \tag{15}$$

Here, $x$ is the full-dose CT image (reference), $y$ indicates the low-dose (or filtered) CT image, and $\mu_x$ and $\mu_y$ are the mean intensity values of the $x$ and y images, respectively. Similarly, $\sigma_x^2$ and $\sigma_y^2$ are the variance of the $x$ and $y$ images, respectively, and $\sigma_{xy}$ stands for the covariance of $x$ and $y$. Constants $c_1$ and $c_2$ are utilized to avoid division by zero (or very small numbers).

Root Mean Square Error (RMSE) measures the difference between CT values in the noisy and reference CT images.

$$RMSE = \sqrt{MSE} \tag{15}$$

*MSE* was defined in Eq. 14.

Relative Bias *(BAIS)* indicates the percentage of error in the low-dose or filtered CT images with respect to the reference CT images.

$$BIAS = (\frac{y-x}{x}) \times 100 \tag{16}$$

Here, $x$ and y denote the CT values in the full-dose and low-dose CT images, respectively.

One of the major challenges in the application of the post-reconstruction denoising techniques is finding the optimal values for the hyper-parameters to achieve the best image quality. Optimal sets of the hyper-parameters would lead to maximal PSNR and SSIM indices as well as minimum RMSE and BIAS. To optimize the hyper-parameters of the Gaussian, Median, and the bilateral filters, a reasonably wide range of values were set for these parameters, and the ones leading to the maximum PSNR and minimum RMSE were selected for the final implementation of these filters. It should be noted that this procedure was repeated independently for each level of the low-dose CT images. Figure 2 shows a representative example of this optimization procedure for the Gaussian hyper-parameter at the low-dose level of 60%. The BIAS, PSNR, and RMSE metrics were calculated for the whole-body, lung, and whole-body without lung, separately.

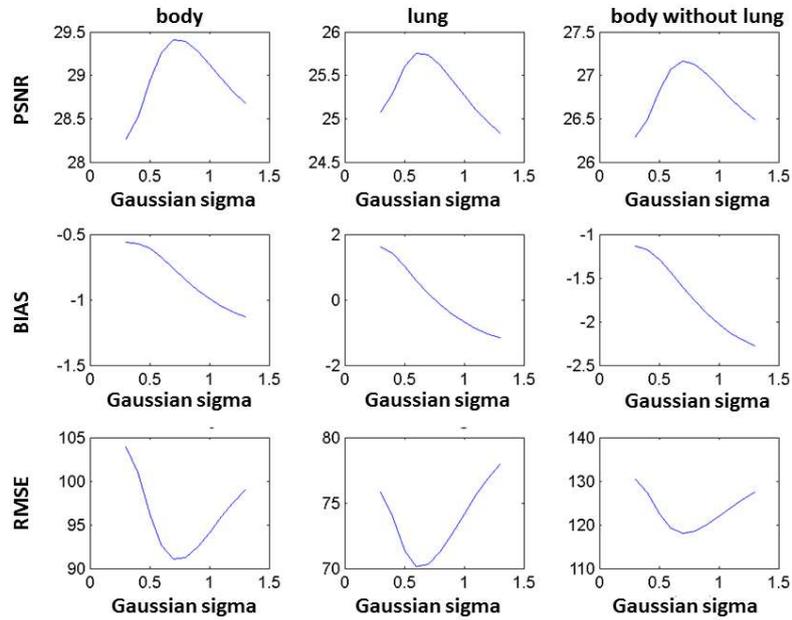

**Fig. 2** The optimal value for the Gaussian hyper-parameter (sigma) at the low-dose level of 60%.

## 3  Results

Figure 3 depicts the simulated low-dose CT images at 60%, 40%, 20%, and 10% of the standard-dose levels for a representative patient with positive Covid-19 test. The quantitative metrics, including SSIM, PSNR, RMSE, and BIAS, calculated for the different low-dose CT images within the whole-body, lung, and whole-body without lung, are reported in Table 1.

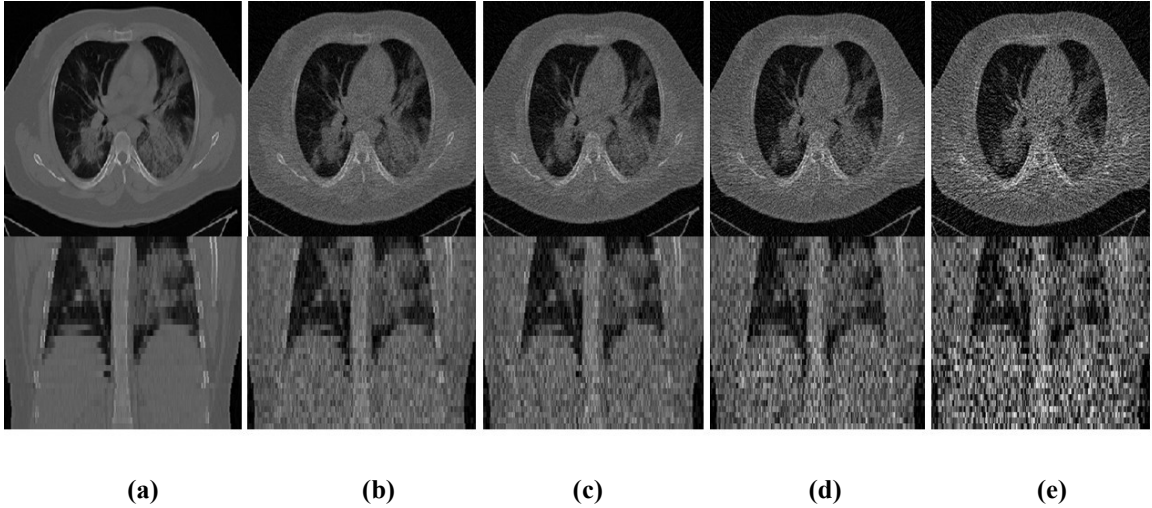

**Fig. 3** Simulated low-dose CT images for a representative patient with positive Covid-19 test. **(a)** Reference CT image (full-dose). **(b)** 60% of full-dose. **(c)** 40% of full-dose. **(d)** 20% of full-dose. **(e)** 10% of full-dose.

**Table 1** Mean±standard deviation of PSNR, SSIM, RMSE, and BIAS metrics calculated for the different low-dose CTs within the whole-body, lung, and whole-body without lung across 49 patients.

| 60% Dose | SSIM | PSNR | RMSE (HU) | BIAS (%) |
|---|---|---|---|---|
| Whole-body | 0.64±0.09 | 26.64±2.14 | 152.91±27.08 | -0.10±0.23 |
| Lung | 0.93±0.02 | 23.57±2.17 | 124.25±27.96 | 1.50±0.69 |
| Whole-body without lung | 0.71±0.09 | 26.17±2.07 | 160.97±25.49 | -0.14±0.25 |
| **40% Dose** | | | | |
| Whole-body | 0.59±0.10 | 24.66±2.20 | 191.80±38.23 | 0.18±0.31 |
| Lung | 0.92±0.02 | 21.95±2.30 | 151.10±37.32 | 1.99±0.93 |
| Whole-body without Lung | 0.68±0.09 | 24.16±2.12 | 202.63±36.64 | 0.14±0.32 |
| **20% Dose** | | | | |
| Whole-body | 0.53±0.11 | 20.90±2.38 | 297.41±68.05 | 1.26±0.71 |
| Lung | 0.90±0.02 | 18.89±2.59 | 217.37±64.66 | 4.30±1.85 |
| Whole-body without lung | 0.63±0.10 | 20.33±2.29 | 316.46±66.43 | 1.17±0.70 |
| **10% Dose** | | | | |
| Whole-body | 0.47±0.11 | 17.22±2.33 | 452.90±94.95 | 4.09±1.75 |
| Lung | 0.88±0.02 | 15.54±2.78 | 322.08±103.99 | 11.44±4.49 |
| Whole-body without lung | 0.58±0.10 | 16.62±2.21 | 483.14±89.84 | 3.78±1.63 |

Table 2 presents the quantitative metrics calculated after the application of the Gaussian filter to the different low-dose CT images. The SSIM and PSNR indices increased after the application of the Gaussian filters, which indicates the overall improved quality of the resulting images. Figure 4 presents the low-dose CT images after the application of the Gaussian filter (for the same views presented in Fig. 2). Visual inspection revealed effective noise reduction; however, some streak-like artifacts are seen in the 10% low-dose image.

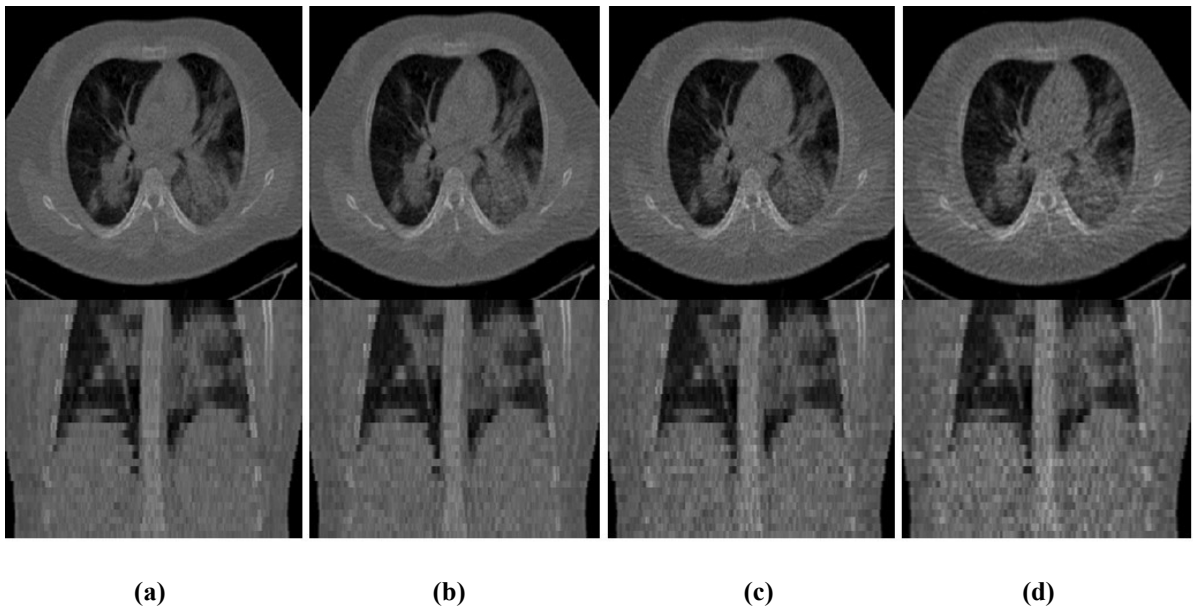

     (a)       (b)       (c)       (d)

**Fig. 4** (a) 60% low-dose, (b) 40% low-dose, (c) 20% low-dose, and (d) 10% low-dose CT images after application of the Gaussian filter.

**Table 2.** Mean±standard deviation of PSNR, SSIM, RMSE, and BIAS calculated for the different low-dose CTs after application of the Gaussian filter within the whole-body, lung, and whole-body without lung across 49 patients.

| **60% Dose** | SSIM | PSNR | RMSE | BIAS |
|---|---|---|---|---|
| Whole-Body | 0.74±0.06 | 30.97±1.84 | 91.30±6.40 | -0.51±0.39 |
| Lung | 0.93±0.01 | 26.62±1.43 | 86.36±9.73 | -1.29±0.52 |
| Whole-Body Without Lung | 0.78±0.09 | 28.97±1.99 | 115.15±10.96 | -1.47±0.62 |
| **40% Dose** | | | | |
| Whole-Body | 0.72±0.07 | 30.13±1.79 | 100.66±8.08 | -0.23±0.47 |
| Lung | 0.93±0.01 | 26.10±1.50 | 91.82±11.77 | -0.80±0.75 |
| Whole-BodyWithout Lung | 0.77±0.06 | 28.34±1.86 | 123.72±9.48 | -1.19±0.70 |
| **20% Dose** | | | | |
| Whole-Body | 0.69±0.08 | 28.61±1.82 | 120.20±12.97 | 0.75±0.90 |
| Lung | 0.93±0.01 | 25.09±1.60 | 103.46±15.70 | 1.02±1.68 |
| Whole-BodyWithout Lung | 0.78±0.07 | 27.04±1.79 | 143.5±10.93 | -0.40±1.12 |
| **10% Dose** | | | | |
| Whole-Body | 0.66±0.08 | 26.51±1.93 | 153.60±21.12 | 3.48±1.96 |
| Lung | 0.92±0.02 | 23.43±1.89 | 126.08±25.10 | 7.70±4.33 |
| Whole-Body without Lung | 0.71±0.07 | 25.27±1.80 | 176.34±15.90 | 2.01±2.07 |

Table 3 summarizes the evaluation parameters calculated for the different low-dose CT images after the application of the median filter. Compared to the Gaussian filter, the median filter led to noticeable signal loss, particularly in 10% low-dose CT images with more than -9% bias in the lung region. Overall, the median filter exhibited inferior performance compared to the Gaussian filter with increased RMSE and reduced SSIM. Figure 5 displays the median filtered low-dose CT images wherein less effective noise reduction is observed compared to the Gaussian filtered images in Fig. 4.

**Table 3.** Mean±standard deviation of PSNR, SSIM, RMSE, and BIAS were calculated for the different low-dose CTs after application of the median filter within the whole-body, lung, and whole-body without lung across 49 patients.

| **60% Dose** | SSIM | PSNR | RMSE | BIAS |
|---|---|---|---|---|
| Whole-Body | 0.71±0.07 | 30.79±1.79 | 93.27±7.26 | -0.60±0.37 |
| Lung | 0.93±0.01 | 26.21±1.50 | 90.44±9.60 | -1.99±0.77 |
| Whole-Body without Lung | 0.77±0.07 | 29.94±1.81 | 102.85±8.55 | -0.75±0.38 |
| **40% Dose** | | | | |
| Whole-Body | 0.67±0.08 | 29.70±1.81 | 105.96±10.86 | -0.43±0.44 |
| Lung | 0.93±0.02 | 25.36±1.61 | 99.98±13.11 | -2.32±0.96 |
| Whole-Body without Lung | 0.74±0.07 | 28.87±1.78 | 116.41±10.75 | -0.55±0.44 |
| **20% Dose** | | | | |
| Whole-Body | 0.63±0.09 | 28.09±1.81 | 127.42±11.55 | -0.28±0.80 |
| Lung | 0.91±0.03 | 23.12±1.62 | 129.60±18.09 | -6.15±2.24 |
| Whole-Body without Lung | 0.72±0.09 | 25.52±2.09 | 171.97±21.43 | -1.19±1.81 |
| **10% Dose** | | | | |
| Whole-Body | 0.55±0.10 | 25.92±1.88 | 163.86±18.02 | 1.00±1.40 |
| Lung | 0.86±0.04 | 21.61±1.85 | 155.10±27.90 | -9.16±4.02 |
| Whole-Body without Lung | 0.66±0.09 | 23.47±2.04 | 217.63±25.04 | 0.04±1.69 |

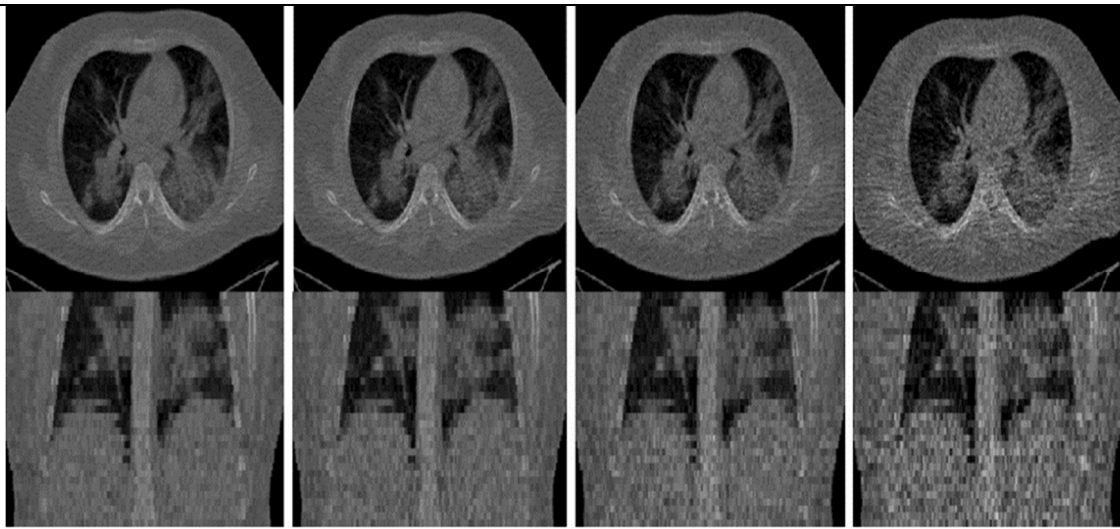

    **(a)**      **(b)**      **(c)**      **(d)**

**Fig. 5** (a) 60% low-dose, (b) 40% low-dose, (c) 20% low-dose, and (d) 10% low-dose CT images after application of the median filter.

Similarly, the quantitative evaluation of the Bilateral filtering in terms of PSNR, SSIM, RMSE, and BIAS parameters are reported in table 4, wherein reduced quantitative bias and improved SSIM and PSNR are observed compared to the Gaussian and median filters. The Bilateral filtering led to a maximum bias of less than 5% compared to the Gaussian and median filters with a bias of more than 7% and -9%, respectively. Figure 6 illustrates the effective noise reduction in the Bilateral filtered images, particularly in the 20% low-dose CT image wherein the pattern of infection can be clearly seen in the lung region with minimal signal loss and, or artifact.

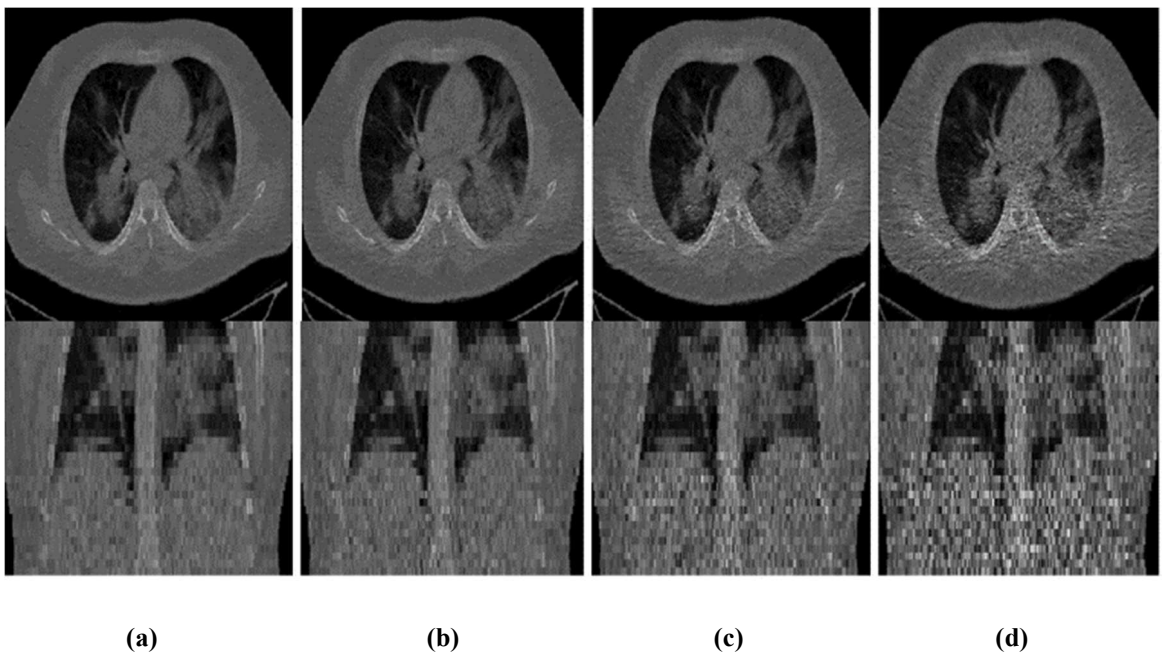

(a)  (b)  (c)  (d)

**Fig. 6** (a) 60% low-dose, (b) 40% low-dose, (c) 20% low-dose, and (d) 10% low-dose CT images after application of the Bilateral filter.

Table 4. Mean±standard deviation of PSNR, SSIM, RMSE, and BIAS calculated for the different low-dose CTs after application of the Bilateral filter within the whole-body, lung, and whole-body without lung across 49 patients.

| 60% Dose | SSIM | PSNR | RMSE | BIAS |
|---|---|---|---|---|
| Whole-Body | 0.73±0.09 | 31.75±1.80 | 83.54±7.31 | -0.36±0.30 |
| Lung | 0.93±0.01 | 27.45±1.90 | 79.23±14.84 | 0.20±0.31 |
| Whole-BodyWithout Lung | 0.80±0.08 | 31.75±1.83 | 82.27±8.66 | -0.25±0.26 |
| **40% Dose** | | | | |
| Whole-Body | 0.70±0.10 | 30.80±1.83 | 93.22±7.65 | -0.33±0.46 |
| Lung | 0.92±0.02 | 26.92±1.71 | 83.95±13.61 | -0.56±0.55 |
| Whole-BodyWithout Lung | 0.75±0.09 | 30.51±1.82 | 96.46±9.05 | -0.54±0.45 |
| **20% Dose** | | | | |
| Whole-Body | 0.61±0.10 | 28.85±1.85 | 116.82±11.52 | 0.40±0.90 |
| Lung | 0.91±0.02 | 25.38±1.68 | 100.21±16.47 | -0.21±1.20 |
| Whole-BodyWithout Lung | 0.71±0.10 | 27.73±1.88 | 132.94±12.93 | -0.47±0.99 |
| **10% Dose** | | | | |
| Whole-Body | 0.54±0.11 | 26.41±2.01 | 160.79±27.35 | 2.83±1.75 |
| Lung | 0.89±0.02 | 23.41±2.04 | 127.06±28.76 | 4.91±3.21 |
| Whole-BodyWithout Lung | 0.65±0.10 | 25.01±1.91 | 182.41±24.83 | 1.54±1.81 |

To compare the performance of the different denoising techniques for the different low-dose levels, boxplots of the RMSE and BIAS parameters are presented in Fig. 7. It is clearly seen that the Bilateral filter resulted in noticeably lower BIAS and RMSE in the four low-dose CT images within the lung region. Although the Bilateral filter enabled remarkable noise reduction, the BIAS in the 10% low-dose CT images within the lung region is yet significantly high, which might skew the correct diagnosis of the infection in the lung region.

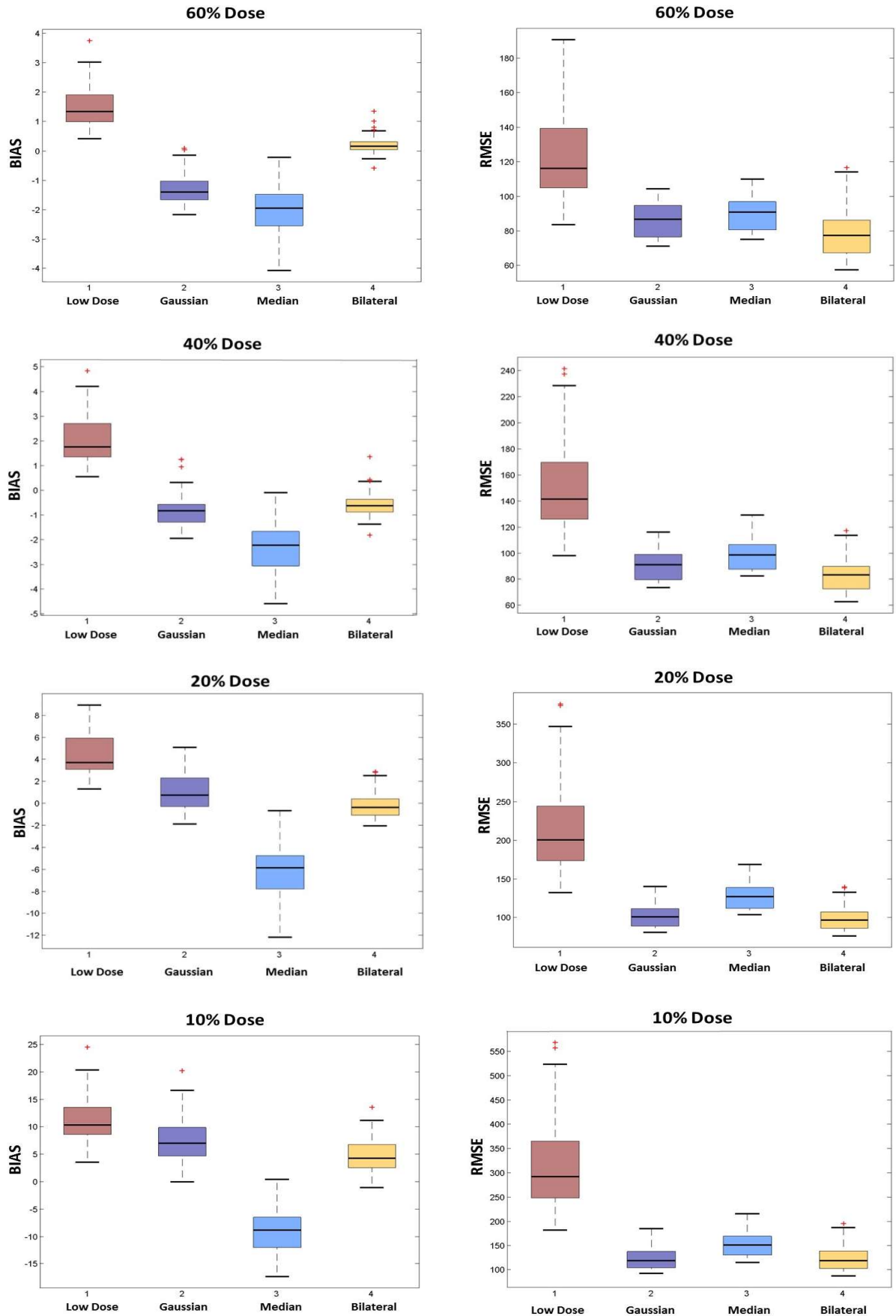

**Fig. 7** Boxplots of BIAS (%) and RMSE (HU) metrics calculated for the different denoising techniques and the different low-dose levels within the lung region.

## 4 Discussion

In this work, four different low-dose CT imaging, namely 10%, 20%, 40%, and 60% of the standard-dose, were simulated for a group of Covid-19 patients. The aim of this study was to investigate to what extent low-dose imaging would deteriorate the quality of CT images for Covid-19 patients and how post-reconstruction filtering could recover the missing signals and underlying structures. Since Covid-19 disease causes infection in the lung, noise levels in this region were separately assessed before and after applying the various denoising techniques.

Regarding Fig. 7, dose reduction in CT imaging up to 60% and 40% of the standard dose would not cause a remarkable noise rise in the lung region, wherein quantitative bias of less than 2% was observed. On the other hand, the noise levels increased dramatically in the 20%- and 10%- low dose images, causing more than 4% and 11% quantitative bias in the lung region, respectively. Although the bilateral filter was able to reduce quantitative bias in the lung region up to -0.21% for 20%-dose and 4.91% for 10%-dose imaging, noticeable streak-like artifacts were observed over the Covid-19 infections within the lung, particularly in 10%-dose images. These image artifacts might skew the accurate diagnosis of Covid-19 and, or identifying or grading the infection in the lung. Thus, 10%-dose imaging, even after the application of the bilateral filter, would not be a reasonable option for the diagnosis of Covid-19 patients.

One of the most important issues in CT imaging of Covid-19 patients is the diagnosis, grading, and classification of the infection in the lung region. In this regard, it is essential to preserve the underlying structures and patterns of the lung region within low-dose imaging and post-reconstruction noise suppression. Hilts et al.[29] concluded that the simple moving average filtering would be able to effectively suppress the noise in low dose CT imaging; however, the noticeable loss of signals and structures in the resulting images, particularly in the lung region, greatly limits its application for Covid-19 patients. Thanh et al.[32], demonstrated that hyper-parameter optimization for the bilateral filter is challenging as the optimal values may vary across the subjects due to different noise levels and, or underlying structures. However, in this study, the lung region was not dedicatedly investigated to relate the findings to the Covid-19 CT imaging.

In this work, the lung segmentation algorithm implemented in the PTK package was used for the delineation of the lung region. Alnaser et al. compared the PTK approach with four different algorithms for the segmentation of the lung, which demonstrated the superior performance of the PTK algorithm.[36] Though the dataset in this study contained challenging lung parenchyma tissues due to the presence of Covid-19 infection, the automated lung segmentation exhibited acceptable accuracy, wherein slight miss-classification errors were manually corrected.

Optimization of the hyper-parameters is the key to the efficient performance of different denoising techniques. To this end, the hyper-parameters of each filter were separately optimized for each dose level. An ideal denoising algorithm would lead to zero quantitative bias and RMSE with respect to the reference images. As such, the hyper-parameters which resulted in minimum RMSE or BIAS, as well as maximum PSNR, were selected dedicatedly for each dose level and denoising technique. It should also be mentioned that the levels of noise presented in the simulated low-dose CT images depend on the CT scanner, acquisition protocol, as well as the resolution of the CT images. Therefore, the optimization of these hyper-parameters should be repeated when the input CT images are from different scanners or acquired through the use of different acquisition protocols.

Overall, considering Fig.7, the Bilateral filter led to a noticeable smaller quantitative bias and RMSE compared to the other denoising techniques, regardless of the noise levels in the input images. The quantitative bias observed in the Bilateral filtered images was less than 5% even in 10%-dose images, which is a clinically tolerable error. However, due to the presence of the streak-like artifacts in the lung, which might skew the correct classification of the Covid-19 infection, 10%-dose imaging is not recommended even after bilateral filtering. On the other hand, 20%-dose CT imaging would offer an acceptable compromise between dose reduction and image quality wherein the quantitative bias in the lung region did not exceed 5% and -1% before and after the application of the bilateral filter.

Regarding the quantitative errors in the lung region and the image artifacts, which might be introduced in the lung region due to the low-dose imaging and, or post-reconstruction denoising, no more than 40% low-dose imaging is recommended if no denoising approach is supposed to be applied. Nevertheless, the application of the bilateral filter to the 20% low-dose CT images led to a similar image quality of the 60% low-dose images without applying any denoising filters. Dose reduction beyond 20% is not recommended for Covid-19 patients as the quantitative bias as well as image artifact would impair the correct diagnosis or grading of Covid-19 disease.

## 5  Conclusion

This work set out to investigate low-dose CT imaging for Covid-19 patients, and in particular, to what extent acquisition dose could be reduced with no significant quantitative bias or image artifact. For this purpose, different low-dose CT imaging, namely 60%, 40%, 20%, and 10% of the standard-dose, were simulated and quantitatively assessed before and after applying the common denoising techniques. Dose reduction beyond 20% introduced considerable quantitative bias as well as image artifact, which may skew the correct diagnosis or grading of Covid-19 infection in the lung region; thus dose reduction beyond 20% is not recommended for Covid-19 patients. The quantitative evaluation of the low-dose CT images indicated the superior performance of the bilateral filter, particularly in the lung region compared to the Gaussian and median filters, which resulted in less than 5% quantitative bias in the lung for the 20%-dose images.